\documentclass[aps,twocolumn,superscriptaddress,english,10pt,nofootinbib,preprintnumbers]{revtex4-2}

\newcommand{\kms}{\rm \,\,km\,s^{-1}Mpc^{-1}}
\newcommand{\fede}{f_{\rm EDE}}
\newcommand{\logzc}{\mathrm{log}_{10}(z_c)}

\newcommand{\eV}{\,\text{eV}}

\newcommand{\Gpc}{\,\text{Gpc}}
\newcommand\hMpc{h\text{Mpc}^{-1}}

\usepackage{float}

\usepackage{amssymb, amsmath, bm, dcolumn, epsf, graphicx, latexsym, slashed}
\usepackage[utf8]{inputenc}

\usepackage{color}

\def\be{\begin{equation}}
\def\ee{\end{equation}}
\def\bea{\begin{eqnarray}}
\def\eea{\end{eqnarray}}

\usepackage{hyperref}
\usepackage{comment}

\usepackage[normalem]{ulem}

\begin{document}

\preprint{INR-TH-2020-041}

\title{Exploring Early Dark Energy solution to the Hubble tension \\with Planck and SPTPol data}

\author{Anton Chudaykin}\email{chudy@ms2.inr.ac.ru}
\affiliation{Department of Physics \& Astronomy, McMaster University,\\ 
1280 Main Street West, Hamilton, ON L8S 4M1, Canada}
\affiliation{Institute for Nuclear Research of the	Russian Academy of Sciences, \\ 
60th October Anniversary Prospect, 7a, 117312
Moscow, Russia}
\author{Dmitry Gorbunov}\email{gorby@ms2.inr.ac.ru}
\affiliation{Institute for Nuclear Research of the	Russian Academy of Sciences, \\ 
60th October Anniversary Prospect, 7a, 117312
Moscow, Russia}
\affiliation{Moscow Institute of Physics and Technology,\\
	Institutsky lane 9, Dolgoprudny, Moscow region, 141700, Russia}
\author{Nikita Nedelko}\email{nikita.nedelko1999@gmail.com}
\affiliation{Institute for Nuclear Research of the	Russian Academy of Sciences, \\ 
	60th October Anniversary Prospect, 7a, 117312
	Moscow, Russia}


\begin{abstract} 
A promising idea to resolve the long standing Hubble tension is to postulate a new subdominant dark-energy-like component in the pre-recombination Universe which is traditionally termed as the Early Dark Energy (EDE). However, as shown in Refs. \cite{Hill:2020osr,Ivanov:2020ril} the cosmic microwave background (CMB) and large-scale structure (LSS) data impose tight constraints on this proposal. Here, we revisit these strong bounds considering the Planck CMB temperature anisotropy data at large angular scales and the SPTPol polarization and lensing measurements. As advocated in Ref. \cite{Chudaykin:2020acu}, this combined data approach predicts the CMB lensing effect consistent with the $\Lambda$CDM expectation and allows one to efficiently probe both large and small angular scales. Combining Planck and SPTPol CMB data with the full-shape BOSS likelihood and information from photometric LSS surveys in the EDE analysis we found for the Hubble constant $H_0=69.79\pm0.99\kms$ and for the EDE fraction $\fede<0.094\,(2\sigma)$. These bounds obtained without including a local distance ladder measurement of $H_0$ (SH0ES) alleviate the Hubble tension to a $2.5\sigma$ level. Including further the SH0ES data we obtain $H_0=71.81\pm1.19\kms$ and $\fede=0.088\pm0.034$ in full accordance with SH0ES. We also found that a higher value of $H_0$ does not significantly deteriorate the fit to the LSS data. Overall, the EDE scenario is (though weakly) favoured over $\Lambda$CDM even after accounting for unconstrained directions in the cosmological parameter space.
We conclude that the large-scale Planck temperature and SPTPol polarization measurements along with LSS data do not rule out the EDE model as a resolution of the Hubble tension. This paper underlines the importance of the CMB lensing effect for robust constraints on the EDE scenario. 


\end{abstract}

\maketitle

\section{Introduction}
\label{sec:intro}

The precise determination of the present-day expansion rate of the Universe, expressed by the Hubble constant $H_0$, is one of the most challenging tasks of modern cosmology. 
Indeed, the value of $H_0$ inferred from different observations appears to be in persistent discrepancy which is conventionally treated as a tension between the {\it direct} (local) and {\it indirect} (global) measurements of the Hubble constant. Namely, the Planck measurement of $H_0$ coming from the cosmic microwave background (CMB) \cite{Aghanim:2018eyx} disagrees with the SH0ES result \cite{Riess:2019cxk}, based on traditional distance ladder approach utilizing Type Ia supernova, at $4.4\sigma$ level. The significance of this tension makes it unlikely to be a statistical fluctuation and hence requires an explanation. 

Numerous local or late-time observations can provide an independent cross-check on the Cepheid-based $H_0$ measurements. In particular, the SN luminosity distances can be calibrated by Miras, $H_0=73.3\pm3.9\kms$ \cite{Huang:2019yhh} and the Tip of the Red Giant Branch (TRGB) in the Hertzsprung-Russell diagram, $H_0=69.6\pm1.9\kms$ \cite{Freedman:2020dne}. Alternatively, local measurements can be performed without relying on any distance ladder indicator through very-long-baseline interferometry observations of water megamasers $H_0=73.9\pm3.0\kms$ \cite{Pesce:2020xfe}, using strongly-lensed quasar systems $H_0=73.3^{+1.7}_{-1.8}\kms$ \cite{Wong:2019kwg} 
\footnote{Recently, the modeling error in time delay cosmography under the assumptions on the form of the mass density profile has been questioned \cite{Kochanek:2019ruu,Blum:2020mgu}. 
The thing is that there is a significant mass-sheet degeneracy that leaves the lensing observables unchanged while rescaling the absolute time delay, and thus alters the inferred $H_0$. The common strategy to deal with that is to make assumptions on the mass density profile motivated by local observations as done by H0LiCOW\,\cite{Wong:2019kwg}. An alternative approach is to partially constrain this inherent degeneracy exclusively by the kinematic information of the deflector galaxy that brings much looser constraints, $H_0=74.5^{+5.6}_{-7.1}\kms$ for a sample of 7 lenses (6 from H0LiCOW) and $H_0=67.4^{+4.1}_{-3.2}\kms$ when a set of 33 strong gravitational lenses from the SLACS sample is used \cite{Birrer:2020tax}. This hierarchical analysis fully accounts for the mass-sheet degeneracy in the error budget that statistically validates the mass profile assumptions made by H0LiCOW \cite{Birrer:2020tax}.
} and gravitational wave signal from merging binary neutron stars \cite{Abbott:2017xzu,Palmese:2020aof}. All these measurements affected by completely different possible systematics agree with each other and give persistently higher values of $H_0$ being in conflict with the Planck prediction \cite{Verde:2019ivm}.
The Hubble tension can be explained by the impact of possible systematics in the Planck data. Indeed, it has been found that the Planck data suffer from multiple internal inconsistencies that can potentially obscure the cosmological inference \cite{Ade:2015xua,Addison:2015wyg,Aghanim:2018eyx}. The most significant feature refers to an interesting oscillatory shape in the TT power spectrum that resembles an extra lensing smoothing of the CMB peaks compared to the $\Lambda$CDM expectation. 
The significance of this "lensing anomaly" is rather high, $2.8\sigma$ \cite{Motloch:2019gux}, while no systematics in the Planck data has been identified so far \cite{Aghanim:2016sns,Aghanim:2018eyx}. Such inconsistencies force one to consider independent measurements of the CMB anisotropies, especially on small scales. Ground based observations provided by the South Pole Telescope (SPT) \cite{Story:2012wx,Henning:2017nuy} and the Atacama Cosmology Telescope (ACT) \cite{Aiola:2020azj,Choi:2020ccd} perfectly suit for this purpose since they probe exclusively small angular scales. These observations indicate internally consistent gravitational lensing of CMB, i.e. the lensing information deduced from the smoothing of acoustic peaks at high-$\ell$ agrees  well with the predictions of 'unlensed' CMB temperature and polarization power spectra \cite{Henning:2017nuy,Han:2020qbp}.

A more beneficial approach is to combine ground based observations with the full sky surveys. Indeed, ground-based telescopes are sensitive to much smaller angular scales unattainable in full sky surveys that can bring a noticeable cosmological gain. Recently, combined data analysis based on the Planck temperature and SPTPol polarization and lensing measurements found a substantially higher value $H_0=69.68\pm1.00\kms$ \cite{Chudaykin:2020acu} that alleviates the Hubble tension to $2.5\sigma$ statistical significance within the $\Lambda$CDM cosmology. It also completely mitigates the so-called $S_8$ tension between different probes of Large Scale Structure (LSS) statistics and the Planck measurements \cite{DiValentino:2020vvd}.
It implies that the mild tension between the LSS and CMB data is solely driven by an excess of the lensing-induced smoothing of acoustic peaks observed in the Planck temperature power spectrum at high-$\ell$ \cite{Chudaykin:2020acu,Addison:2015wyg,Aghanim:2016sns}.


Besides that, the information about the present-day expansion rate of the Universe can be extracted from different measurements at low redshifts calibrated by any early-universe data independently of any CMB data. This is done by combining LSS observations with primordial deuterium abundance measurements. First such measurement comes from the baryon acoustic oscillation (BAO) experiments. Utilizing the BAO data from the Baryon Oscillation Spectroscopic Survey (BOSS) \cite{Alam:2016hwk}, the prior on $\omega_b$ inferred from the Big Bang Nucleosynthesis (BBN) \cite{Cooke:2016rky} and late-time probe of the matter density from the Dark Energy Survey (DES) \cite{Abbott:2017wau} yields $H_0 = 67.4^{+1.1}_{-1.2}\kms$ \cite{Abbott:2017smn}. 
Measurements of BAO scales for galaxies and the Ly$\alpha$ forest \cite{Blomqvist:2019rah} augmented with the BBN prior bring similar estimate $H_0 =67.6\pm1.1\kms$ \cite{Aubourg:2014yra,Cuceu:2019for,Schoneberg:2019wmt}. Second, the Hubble constant measurement can be accomplished with galaxy clustering alone using the full-shape (FS) information of the galaxy power spectrum \cite{Ivanov:2019pdj,DAmico:2019fhj,Philcox:2020vvt}. In particular, the joint FS+BAO analysis brings $H_0 = 68.6 \pm 1.1\kms$ \cite{Philcox:2020vvt}.

Importantly, all these measurements assume the standard evolution of the Universe prior to recombination. It sticks the sound horizon $r_s$ to the $\Lambda$CDM function of cosmological parameters. However, any sizable shift in $H_0$ value that needed to solve the Hubble tension must be accompanied by corresponding modification of $r_s$ to preserve the fit to CMB data that measure the angular scale of the sound horizon $\theta_s$ with a very high accuracy. This modification can be accomplished by introducing a new component which increases $H(z)$ in the decade of scale factor evolution prior to recombination. Such early-universe scenarios have been advocated as the most likely solution of the Hubble tension in Ref. \cite{Knox:2019rjx}. The broad subclass of these models has been termed Early Dark Energy (EDE). Many EDE-like scenarios have been proposed from a phenomenological point of view \cite{Kamionkowski:2014zda,Poulin:2018a,Poulin:2018cxd,Smith:2019ihp,Agrawal:2019lmo,Lin:2019qug,Ye:2020btb,Niedermann:2020qbw}, whilst others present concrete realizations of particle-physics models \cite{Niedermann:2019olb,Alexander:2019rsc,Kaloper:2019lpl,Berghaus:2019cls}. It is pertinent to highlight two interesting realizations \cite{Sakstein:2019fmf,Braglia:2020bym} in which the EDE field becomes dynamical precisely around matter-radiation equality that ameliorates the coincidence problem inherent to most EDE implementations.

We examine one popular EDE implementation which postulates a dynamical scalar field which behaves like dark energy in early times and then rapidly decays in a relatively narrow time interval near matter-radiation equality. The increased energy density of the Universe prior to recombination shrinks the comoving sound horizon $r_s$ which lifts up $H_0$ while keeping the angular scale $\theta_s$ intact. This extension of the $\Lambda$CDM model can be parameterized by 3 parameters: the maximal injected EDE fraction $\fede$, the critical redshift $z_c$ at which this maximum is reached and an initial scalar field value denoted by dimensional quantity $\theta_i$ (in analogy to the axion misalignment angle \cite{Dine:1982ah,Abbott:1982af,Preskill:1982cy}). It has been previously established that this prescription allows for values of $H_0$ consistent with SH0ES whilst preserving the fit to the CMB data \cite{Poulin:2018cxd,Smith:2019ihp}.

The situation becomes more intricate when LSS data are taken into account. The thing is that the EDE scenario matches the CMB data at the cost of shifting several cosmological parameters that is not compatible with LSS data. In particular, it substantially increases the physical density of cold dark matter $\omega_c$ and to a lesser extent the spectral index $n_s$ that raise up the late-time parameter $S_8=\sigma_8\sqrt{\Omega_m/0.3}$. This change  exacerbates the $S_8$ tension between LSS observables and the Planck data and imposes tight constraints on the EDE scenario as a possible solution to the Hubble tension \cite{Hill:2020osr}. Namely, when considering all LSS data with the Planck, SNIa, BAO and RSD measurements one finds $\fede<0.06\,(2\sigma)$ \cite{Hill:2020osr} which is well below the value needed to resolve the Hubble tension, $\fede\sim0.1$. The main driver of this strong constraint is the overly enhanced lensing-induced smoothing effect that affects the Planck temperature power spectrum at high-$\ell$ and pulls the late-time amplitude to a higher value \cite{Addison:2015wyg,Aghanim:2016sns} being in conflict with the LSS data. 
It has been shown that the tension between the Planck and various LSS probes can be reconciled if one combines the large-angular scale Planck temperature measurements \footnote{The idea of separating large and small angular scales in the Planck temperature power spectrum has been thoroughly investigated in literature, see e.g.\,\cite{Aghanim:2016sns,Burenin:2018nuf}.} with the ground-based observations of the SPTPol survey as argued in Ref. \cite{Chudaykin:2020acu}. Thus, one expects that the tight LSS constraints on EDE can be alleviated if one replaces the Planck CMB data at high multipoles $\ell$ with the SPTPol measurements. Revising the constraining power of LSS in the EDE model using the different CMB setup that predicts the consistent CMB lensing effect is one of the main goals of this paper.

Another important ingredient of our study is the full-shape analysis of galaxy power spectrum. This treatment is based on complete cosmological perturbation theory with a major input from the Effective Field Theory (EFT) of LSS. This approach includes all necessary ingredients (UV counterterms and IR resummation) needed to reliably describe galaxy clustering on mildly nonlinear scales. The full-shape template of the galaxy power spectrum contains a large amount of cosmological information that can effectively constrain various extensions of the $\Lambda$CDM model. In particular, it has been shown that the full-shape BOSS likelihood yields a $\approx20\%$ improvement on the EDE constraint from the CMB data alone \cite{Ivanov:2020ril}. Crucially, the standard BOSS likelihood does not appreciably shrink the Planck limits due to the lack of full-shape information therein \cite{Ivanov:2020ril}. In order to obtain more refined constraints on EDE, we employ the full-shape BOSS likelihood in our analysis.

In this paper, we examine the EDE scenario using the Planck and SPTPol measurements of the CMB anisotropy. Namely, we follow the combined data approach validated in Ref. \cite{Chudaykin:2020acu} and combine the Planck temperature power spectrum at large angular scales with polarization and lensing measurements from the SPTPol survey \cite{Henning:2017nuy}. This approach ensures the internally consistent CMB lensing effect and allows one to gain cosmological information from both large and small angular scales.

We improve our previous analysis \cite{Chudaykin:2020acu} in several directions. First, we solve the evolution of the scalar field perturbations directly using the Klein-Gordon equation which does not rely on the effective fluid description. Second, we consider a more realistic EDE setup which generalizes a pure power-law potential considered in Ref. \cite{Chudaykin:2020acu}. Third, we use the full BOSS galaxy power spectrum likelihood that yields much stronger constraints on EDE compared to the standard BOSS likelihood \cite{Ivanov:2020ril}.  
Finally, we exploit the more recent LSS data coming from the DES-Y1 \cite{Abbott:2017wau}, Kilo-Degree Survey (KiDS) \cite{Asgari:2020wuj} and Subaru Hyper Suprime-Cam (HSC) \cite{Hikage:2018qbn} measurements that allow us to reduce by half the error bars on $S_8$ compared to that examined in Ref. \cite{Chudaykin:2020acu}.

The outline of this paper is as follows. In Sec. \ref{sec:theory} we review the physics of the EDE scenario. In Sec. \ref{sec:constr} we present the combined data approach, data sets and main results. Finishing in Sec. \ref{sec:conc} we highlight the differences between our approach and previous EDE analyses, interpret our outcomes and discuss the prospects. 

\section{The early dark energy model}
\label{sec:theory}

The main goal of EDE proposal is to decrease the comoving sound horizon of the last scattering epoch, 
\begin{equation}
\label{rs}
r_s(z_*) = \int _{z_*} ^\infty \frac{{\rm d} z}{H(z)} c_s(z) ,
\end{equation}
where $z_*$ denotes the redshift of the last scattering in such a way that the higher value of $H_0$ encoded in the comoving angular diameter distance
\begin{equation}
\label{DA}
D_A(z_*) = \int _0 ^{z_*} \frac{{\rm d} z}{H(z)} ,
\end{equation}
can be accommodated without changing the angular scale of the sound horizon,
\begin{equation}
\label{thetas}
\theta_* = \frac{r_s (z_*)}{D_A(z_*)}.
\end{equation}

Necessary adjustments of the early-universe dynamics can be readily understood. The angular diameter distance defined by \eqref{DA} is driven by the low-redshift cosmic evolution and, hence, directly relies on $H_0$. Eq. \eqref{thetas} implies that the upward shift in $H_0$ must be accompanied by the downward shift of $r_s$ since $\theta_*$ is measured to $0.03\%$ precision by Planck. However, the sound horizon given by \eqref{rs} is saturated near the lower bound of the integral that requires the increased expansion rate of the Universe at times shortly before recombination.
In EDE scenarios such increase is provided by an additional contribution to the total energy density of the Universe which acts as dark energy at early times. The magnitude of the Hubble tension dictates the energy scale of the early-time contribution to be of order $\sim\eV$. Crucially, this extra energy density initially stored in EDE must rapidly decay and practically disappear before the last scattering so as not to affect the CMB anisotropy on small scales.

The simplest model where the requisite dynamics can be realized is that of the scalar field. Indeed, in this scenario at high redshifts the scalar field is frozen and acts as EDE whereas afterwards it begins to oscillate and its energy density redshifts like matter density, $\rho \propto a^{-3}$. In the context of particle physics, the candidate for this scalar field can be the axion \cite{Peccei:1977hh,Wilczek:1977pj,Weinberg:1977ma} with a periodic potential $V \propto m^2 f^2 \cos \phi/f$ generated by non-perturbative effects. However, the EDE field must decay as radiation or faster to keep the late-time evolution intact, while in the simplest example of axion-like model the EDE energy density redshifts as matter. This obstacle can be overcome in various extensions of the axion physics, see \cite{Kamionkowski:2014zda,Montero:2015ofa}.

We consider a general scalar field endowed with the discrete periodic symmetry and its scalar potential has a generic form $V=\sum c_n\cos(n\phi/f)$. 
Then, we tune the first $n$ coefficients and neglect higher order harmonics to end up with scalar potential of specific form
\begin{equation}
\label{PoulinEDE}
V = V_0 \left( 1 - \cos (\phi/f)\right)^n \;\; , \;\; V_0 \equiv m^2 f^2 \, .
\end{equation}
It has been shown that this type of potential can alleviate the Hubble tension \cite{Poulin:2018cxd}. One observes, that after onset of scalar field oscillations, this potential with $n=2$ affords the dilution of the energy density initially stored in EDE as radiation ($\propto a^{-4}$), and for $n\rightarrow \infty$ it redshifts as  kinetic energy ($\propto a^{-6}$) thereby reproducing the Acoustic Dark Energy scenario \cite{Lin:2019qug}. Recent investigations of the EDE dynamics with potential \eqref{PoulinEDE} in the context of the Hubble tension reveal that the case $n=3$ provides a somewhat better fit to the overall cosmological data \cite{Smith:2019ihp,Agrawal:2019lmo,Chudaykin:2020acu}. 
Potential \eqref{PoulinEDE} can appear with specially tuned model parameters e.g. as a low-energy limit of multi-axion models, where several QCD-like sectors fall into confinement and form a complicated periodic potential for the lightest axion. With specially tuned model parameters the effective potential can take the form \eqref{PoulinEDE}. This construction has been suggested to improve the model of natural inflation, see e.g.\,\cite{Kim:2004rp,Agrawal:2018mkd}. Anyway, it is worth noting that much later, after recombination, hence for much smaller scalar field $\phi$, the effective axion potential may differ, e.g. approach more standard case of $n=1$. The field decays slower then, like scalar dark matter, but has no recognizable impact on late-time cosmology, and so we stick to \eqref{PoulinEDE}. To highlight that our setup is different from the Peccei--Quinn axion, we refer to the scalar field $\phi$ as axion-like particle (ALP) in what follows.

The cosmological dynamics of EDE field with potential \eqref{PoulinEDE} can be succinctly described by the following effective parameters: the maximal fraction of the EDE field in the total energy density of the Universe, $\fede$, and the redshift, $z_c$, at which this maximum is reached. These parameters absorb the particle physics parameters, the ALP mass $m$ and decay constant $f$, and have clear cosmological implication. To solve the Klein-Gordon equation one must also specify the initial field displacement $\theta_i\equiv\phi_i/f$ which represents re-normalized field variable at very early times. There the EDE field remains almost constant, so we start its evolution with zero velocity, and the system quickly approach the slow-roll regime for the EDE field. Later this regime terminates and the field starts non-harmonic oscillations, which frequency is determined by both $\theta_i$ and $m$\,\cite{Smith:2019ihp}. These parameters also govern the perturbed dynamics of the EDE field \cite{Smith:2019ihp}. The last parameter $n$ determines the rate at which the EDE field dilutes. All in all, the EDE dynamics is entirely described by the following four parameters $\fede,\,z_c,\,\theta_i,\,n$.



\section{Constraints on the EDE scenario}
\label{sec:constr}

Parameter estimates presented in this paper are obtained with the combined Einstein-Boltzmann code comprised of \texttt{CLASS\_EDE} \cite{Hill:2020osr} and \texttt{CLASS-PT} \cite{Chudaykin:2020aoj} (both extensions of \texttt{CLASS}~\cite{Blas:2011rf}) \footnote{The code that combines both \texttt{CLASS\_EDE} and  \texttt{CLASS-PT} extensions is available on the web-page \href{https://github.com/Michalychforever/EDE_class_pt}{https://github.com/Michalychforever/EDE\_class\_pt}.}, interfaced with the \texttt{Montepython} Monte Carlo sampler \cite{Audren:2012wb,Brinckmann:2018cvx}. \texttt{CLASS\_EDE} implements both the background and  perturbed dynamics of the scalar field. Namely, it directly solves the homogeneous and perturbed  Klein-Gordon equation along the lines of Ref. \cite{Agrawal:2019lmo,Smith:2019ihp}. We apply adiabatic initial conditions for the scalar field fluctuations in full accordance with \cite{Smith:2019ihp}. We perform the Markov Chain Monte Carlo approach to sample the posterior distribution adopting a Gelman-Rubin \cite{Gelman:1992zz} convergence criterion $R-1 < 0.15$. The plots with posterior densities and marginalized limits are generated with the latest version of the \texttt{getdist} package\footnote{\href{https://getdist.readthedocs.io/en/latest/}{
\textcolor{blue}{https://getdist.readthedocs.io/en/latest/}}
}~\cite{Lewis:2019xzd}. 

Following previous EDE analysis \cite{Hill:2020osr,Ivanov:2020ril} we impose uniform priors on the EDE parameters: $\fede=[0.001,0.5]$, $\logzc=[3,4.3]$ and $\theta_i=[0.1,3.1]$. We translate the effective parameters $\fede$ and $z_c$, to the particle physics parameters, $f$ and $m$, given some initial field displacement $\theta_i$ via a shooting algorithm realized in \texttt{CLASS\_EDE}. We fix $n=3$ as the cosmological data only weakly constrain this parameter \cite{Smith:2019ihp}. We also vary 6 standard $\Lambda$CDM parameters within broad uniform priors: $\omega_c=\Omega_ch^2$, $\omega_b=\Omega_bh^2$, $H_0$, $\ln(10^{10} A_\mathrm{s})$, $n_s$ and $\tau$. We assume the normal neutrino hierarchy with the total active neutrino mass $\sum m_\nu=0.06\eV$. When the full-shape BOSS likelihood is included, all matter transfer functions are calculated along the lines of the standard cosmological perturbation theory that consistently predict galaxy/matter clustering on mildly nonlinear scales. Otherwise, we use the Halofit module to compute nonlinear matter power spectrum which allows us to reliably predict the lensed CMB power spectra and lensing potential power spectrum at high multipoles.

One comment on the choice of EDE priors is in order here. Uniform priors on the EDE parameters $\fede$, $z_c$ and $\theta_i$ implies strongly non-uniform probability distributions for the particle physical parameters, namely the ALP mass $m$ and decay constant $f$. In particular, the posterior distribution for $f$ is peaked near the Planck mass scale \cite{Hill:2020osr}. This drastic departure from a uniform distribution raises the question regarding the dependence of the EDE posteriors on the choice of the flat priors. The analysis of Ref. \cite{Hill:2020osr} demonstrates that uniform priors imposed directly on the particle physics parameters, $f$ and $\log m$, strongly downweight large $\fede$ values, in comparison to uniform priors placed on the effective EDE parameters, $\fede$ and $z_c$. Thereby, the analysis with the flat physical priors further tightens the upper limits on $\fede$ and thus significantly weakens the possibility of resolving the $H_0$ tension. Although imposing flat physical priors is arguably more physically reasonable, we exploit the uniform priors on the effective parameters, $\fede$ and $z_c$, for two reasons. First, the effective parameters have clear cosmological implication that provides direct comparison with other early-universe scenarios, e.g. \cite{Lin:2020jcb,Murgia:2020ryi}. Second, the effective description allows one to cover other physical realizations of the EDE model which have different physical priors, see Ref. \cite{Berghaus:2019cls,Sakstein:2019fmf,Braglia:2020bym,Niedermann:2019olb,Alexander:2019rsc}.

\subsection{Methodology}
\label{subsec:method}

The cornerstone of our analysis is the combined data approach that allows one to extract reliable cosmological information from multiple CMB experiments in a wide range of angular scales, see \cite{Chudaykin:2020acu}. We examine the $\Lambda$CDM and EDE predictions utilizing the Planck large-angular scale measurements along with the ground-based observations of the 500 deg$^2$ SPTPol survey. Before going to the data sets we assert our CMB setup and reveal its importance in the light of the Planck lensing tension.

We combine the Planck TT power spectrum at $\ell<1000$ with the SPTPol measurements of TE, EE spectra following the CMB specification adopted in Ref. \cite{Chudaykin:2020acu}. We do not include the Planck polarisation measurements at intermediate angular scales because the Planck TE and EE spectra have residuals at $\ell\sim30-1000$ relative to the $\Lambda$CDM prediction \cite{Smith:2019ihp}. Given this range of multipoles roughly corresponds to the modes that enter the horizon while the EDE density is important, the Planck polarization measurements strongly disfavour the EDE solution as shown in Ref. \cite{Lin:2020jcb}. Interestingly, the ACT observations do not detect any features in the TE and EE measurements in this multipole region \cite{Aiola:2020azj}. This data discrepancy motivates us to take the TE and EE power spectra entirely from the SPTPol survey which do not manifest any significant residuals relative to $\Lambda$CDM \cite{Henning:2017nuy}.

We further include the SPTPol measurement of the lensing potential power spectrum $C_\ell^{\phi\phi}$. Despite a somewhat higher constraining power of the Planck lensing measurements, we do not include it for the following reasons. First, it allows us to investigate the lensing information entirely encoded in SPTPol maps. Second, despite the fact that the Planck lensing amplitude extracted from quadratic estimators of T -, E- or B-fields is in excellent agreement with that from the Planck power spectra, the former is in a mild tension with the large-angular scale Planck TT and SPTPol data which prefer substantially lower values of fluctuation amplitudes \cite{Aghanim:2016sns,Henning:2017nuy}. To provide a self-consistent cosmological inference we employ the direct measurement of $C_\ell^{\phi\phi}$ from the SPTPol survey that agrees well with SPTPol measurements of TE and EE spectra \cite{Bianchini:2019vxp}. 

\begin{figure}[h]
	\begin{center}
	\hspace{-0.3cm}	\includegraphics[width=0.49\textwidth]{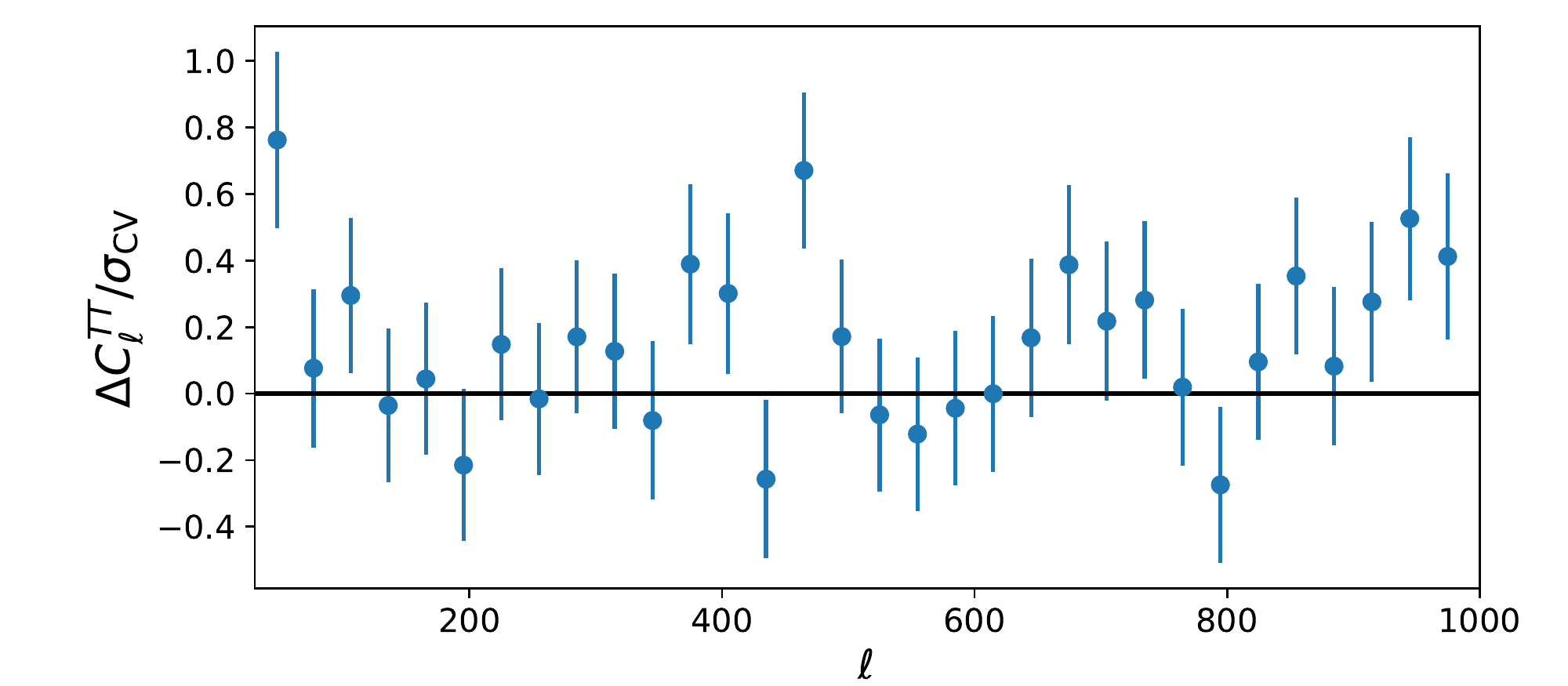}
		\includegraphics[width=0.49\textwidth]{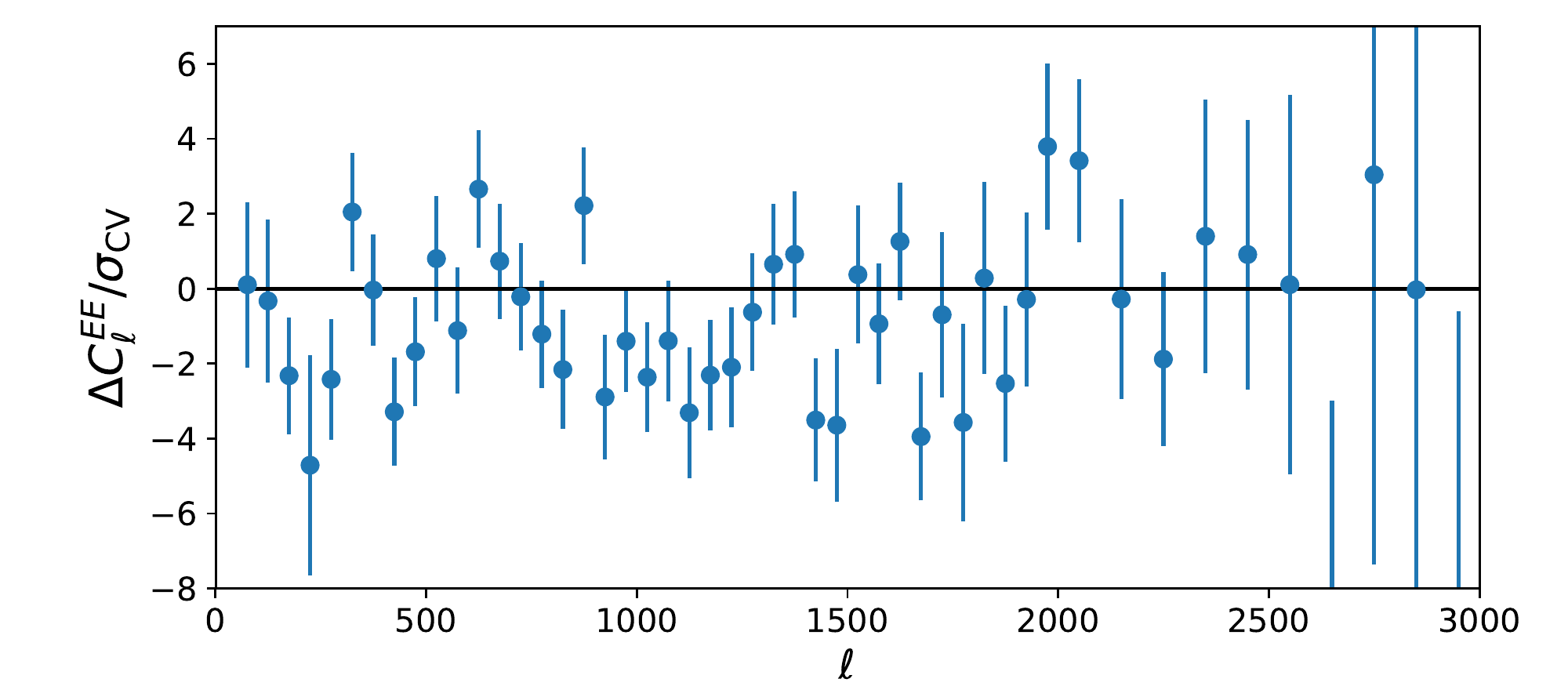}
		\includegraphics[width=0.49\textwidth]{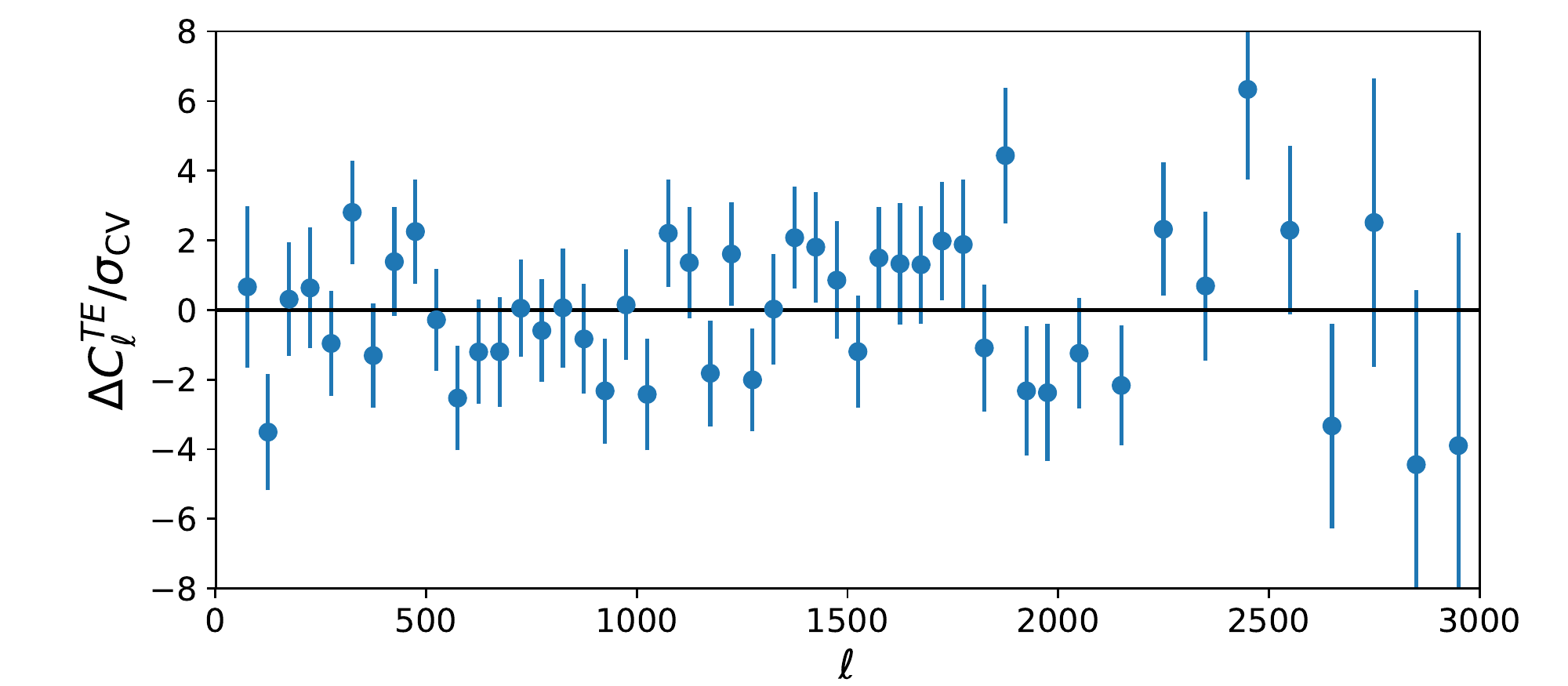}
	\end{center}
	\caption{
		CMB residuals of Planck TT (top panel), SPTPol EE (middle panel) and EE (bottom panel) data with respect to the $\Lambda$CDM model optimized to the $\rm Planck\text{-}low\ell\!+\!SPT$ likelihood.
		\label{fig:residuals} } 
\end{figure}

Statistical agreement between the large-scale Planck temperature and SPTPol polarization measurements has been validated on the level of posterior distributions in Ref. \cite{Chudaykin:2020acu}. Herein, we corroborate this analysis by direct comparison of the spectra to show their consistency in the way that they are used. In Figure \ref{fig:residuals} we show the Planck TT ($30<\ell<1000$) and SPTPol TE and EE ($50<\ell<3000$) residuals relative the $\Lambda$CDM prediction optimized to the $\rm Planck\text{-}low\ell\!+\!SPT$ likelihood \footnote{We do not show the SPTPol polarisazation measurements at high multipoles because the error bars at $3000<\ell<8000$ significantly surpass the cosmic variance.}.
The residuals are shown in units of $\sigma_{\rm CV}$, the cosmic variance error per multipole moment which is given by
\be
\sigma_{\rm CV} =
\begin{cases}
\sqrt{\frac{2}{2\ell+1}} C_\ell^{TT}, & {\rm TT} ,\\
\sqrt{\frac{1}{2\ell+1}}  \sqrt{C_\ell^{TT} C_\ell^{EE} + (C_\ell^{TE})^2}, & {\rm TE} ,\\
\sqrt{\frac{2}{2\ell+1}}  C_\ell^{EE}, & {\rm EE} . \\
\end{cases}
\ee
Once can see that the $\Lambda$CDM predictions match both Planck TT and SPTPol TE and EE measurements well in the range of interest. The corresponding difference between data and theory predictions is comparable to the statistical uncertainties of the Planck and SPTPol data. This analysis paves the way towards direct application of the combined data approach based on the Planck temperature and SPTPol polarization data.

One of the main advantages of the combined data approach presented here is that it predicts a consistent CMB lensing effect within the $\Lambda$CDM cosmology. To make it clear we introduces one phenomenological parameter $A_L$ that scales the lensing potential power spectrum at every point in the parameter space. Utilizing the Planck temperature power spectrum at $\ell<1000$ along with the SPTPol polarization and lensing measurements yields $A_L=0.990\pm0.035$ \cite{Chudaykin:2020acu} being in perfect agreement with the $\Lambda$CDM expectation. In turn, the full Planck data favours overly enhanced lensing-induced smoothing of acoustic peaks which imposes $A_L=1.180\pm0.065$ \cite{Aghanim:2018eyx}. More lensing in the Planck maps may be driven by uncounted systematics in the Planck data on small scales or be a mere statistical fluctuation arising from instrumental noise and sample variance \cite{Aghanim:2016sns,Aghanim:2018eyx}. This effect may be also explained by a proper oscillatory feature in the primordial spectrum generated in specific inflation scenarios \cite{Domenech:2020qay}.
It was recently suggested that the source of the Hubble tension could possibly originate in the modeling of late-time physics within the CMB analysis \cite{Haridasu:2020pms}. Anyway, our combined data approach is free from the Planck lensing anomaly and provides reliable measurements of cosmological parameters.

Noteworthy, the lensing-like anomaly in the Planck data can be partially alleviated by allowing a higher value of the optical depth $\tau$ \cite{Addison:2015wyg}. Indeed, higher $\tau$ gives larger $A_s$ preferred by stronger lensing effect in the CMB maps at high multiploes. However, this would be at odds with the Planck large-scale polarization measurements. In particular, the baseline Planck 2018 measurement of the EE power spectra at $\ell<30$ brings $\tau=0.0506\pm0.0086$ \cite{Aghanim:2018eyx}. Moreover, the improved analysis of the Planck data allows one to reduce the Planck 2018 legacy release uncertainty by $40\%$ which gives $\tau=0.059\pm0.006$ \cite{Pagano:2019tci}. The significantly higher optical depth $\tau=0.089\pm0.014$ was reported by WMAP \cite{Hinshaw:2012aka} but it relies on the old maps of polarized Galactic dust emission \cite{Ade:2015xua}. In our analysis in Fig. \ref{fig:residuals} we follow the conservative Planck 2018 estimate of $\tau$ provided by the $\rm Planck\text{-}low\ell$ likelihood. We argue that even for such a low optical depth, the Planck large-scale temperature and SPTPol polarization measurements predict the consistent CMB lensing effect in the $\Lambda$CDM cosmology \cite{Chudaykin:2020acu} and thus can be used to obtain reasonable cosmological constraints in various extensions of the $\Lambda$CDM model.

\subsection{Data sets}
\label{subsec:data}

We employ the Planck temperature \texttt{Plik} likelihood for multipoles $30\leq\ell<1000$ in the concert with low-$\ell$ \texttt{Commander} TT likelihood, and the low-$\ell$ \texttt{SimAll} EE likelihood \cite{Aghanim:2018eyx}. We vary over all nuisance parameters required to account for observational and instrumental uncertainties. We refer to these measurements as $\rm Planck\text{-}low\ell$.

We utilize CMB polarization measurements from the 500 $\deg^2$ SPTPol survey which includes TE and EE spectra in the multipole range $50<\ell\leq8000$ \cite{Henning:2017nuy}. We vary all necessary nuisance parameters which account for foreground, calibration and beam uncertainties and impose reasonable priors on them. We transform theoretical spectra from unbinned to binned bandpower space using appropriate window functions. We supplement these polarization measurements with the observation of the lensing potential power spectrum  $C_\ell^{\phi\phi}$ in the multipole range $100<\ell<2000$ from the same 500 $\deg^2$ SPTPol survey \cite{Wu:2019hek}. The lensing potential is reconstructed from a minimum-variance quadratic estimator that combines both the temperature and polarization CMB fields. To take into account the effects of the survey geometry we convolve the theoretical prediction for the lensing potential power spectrum with appropriate window functions at each point in the parameter space. We also perturbatively correct $C_\ell^{\phi\phi}$ to address the difference between the recovered lensing spectrum from simulation and the input spectrum along the lines of Ref. \cite{Bianchini:2019vxp}. We denote these SPTPol polarization and lensing measurements \footnote{The complete SPTPol likelihoods for the \texttt{Montepython} environment are publicly available at \href{https://github.com/ksardase/SPTPol-montepython}{https://github.com/ksardase/SPTPol-montepython}.} as SPT in what follows.


We use the data from the final BOSS release DR12 \cite{Alam:2016hwk}, implemented as a joint FS+BAO likelihood, as in Ref. \cite{Philcox:2020vvt}. We refer the reader to \cite{Ivanov:2019pdj,Chudaykin:2019ock} for details of the pre-reconstruction full-shape power spectrum analysis and to Ref. \cite{Philcox:2020vvt} for the strategy of the BAO measurements performed with the post-reconstruction power spectra. Our likelihood includes both pre- and post-reconstruction galaxy power spectrum multipoles $\ell=0,2$ across two different non-overlapping redshift bins with $z_{\rm eff}=0.38$ and $z_{\rm eff}=0.61$ observed in the North and South Galactic Caps (NGC and SGC, respectively). It results in the four different data chunks with the total comoving volume $\simeq6\,(h^{-1}\Gpc)^3$. We create separate sets of nuisance parameters for each data chunk and vary all of them independently. We impose the conservative priors on the nuisance parameters following Ref. \cite{Ivanov:2019pdj}. We present all nuisance parameters and their priors in Appendix \ref{nuisance}. We employ the wavenumber range [$0.01,0.25$] $\hMpc$ for the pre-reconstruction power spectra and [$0.01,0.3$] $\hMpc$ for the BAO measurements based on the post-reconstruction power spectra. We recall this full-shape likelihood as BOSS.

We adopt the SH0ES measurement of the Hubble constant $H_0=74.03\pm1.42\kms$ \cite{Riess:2019cxk}. We impose the Gaussian prior on $H_0$ and call this local measurement as SH0ES.

Finally, we utilize the additional LSS information from multiple photometric surveys. In particular, we consider the DES-Y1 galaxy clustering, galaxy-galaxy lensing and cosmic shear observations \cite{Abbott:2017wau} along with the weak gravitational lensing measurements from KiDS-100 \cite{Asgari:2020wuj} and HSC \cite{Hikage:2018qbn} \footnote{We do not include the cross-correlation between these measurements because the sky overlap between these surveys is small, see \cite{Hill:2020osr}.}. As demonstrated in \cite{Hill:2020osr}, the DES-Y1 "3x2pt" likelihood 
can be well approximated in the EDE analysis by a Gaussian prior on $S_8$. Driven by this observation, we include the DES-Y1, KiDS-100 and HSC measurements via appropriate priors on $S_8$. Namely, for the DES-Y1 combined data analysis we use $S_8=0.773^{+0.026}_{-0.020}$ \cite{Abbott:2017wau}; for the KiDS-100 cosmic shear measurements we adopt $S_8=0.759^{+0.024}_{-0.021}$\cite{Asgari:2020wuj}; for HSC observations we utilize $S_8=0.780^{+0.030}_{-0.033}$ \cite{Hikage:2018qbn}. Finally, we weight each mean with its inverse-variance and obtain the resultant constraint $S_8=0.769\pm0.015$. We include this combined measurement as a Gaussian prior and refer to this simply as $\rm S_8$ in our analysis. 

\subsection{Results}
\label{subsec:spt}

We report our main results obtained from analyses of different cosmological data sets. 

\subsubsection{$Planck\text{-}low\ell\!+\!SPT$}
\label{subsec:cmb}

We firstly examine the cosmological inference from the primary CMB data alone following \cite{Chudaykin:2020acu}. The resulting parameter constraints in the $\Lambda$CDM and EDE models inferred from the $\rm Planck\text{-}low\ell\!+\!SPT$ data are given in Tab. \ref{table:base}. The limits for the $\Lambda$CDM scenario are taken from Ref. \cite{Chudaykin:2020acu} (data set $Base$ therein). The 2d posterior distributions for the EDE model are shown in Fig. \ref{fig:spt_1}.

\begin{table*}[htb!]
	\renewcommand{\arraystretch}{1.1}
	Constraints from $\rm Planck\text{-}low\ell\!+\!SPT$ \vspace{2pt} \\
	\centering
	\begin{tabular}{|l|c|c|}
		\hline Parameter &$\Lambda$CDM~~&~~~EDE \\  \hline\hline
		
		{$\ln(10^{10} A_\mathrm{s})$} & 
		$3.021 \pm 0.017$ &
		$3.024 \pm 0.018$ \\
		
		{$n_\mathrm{s}$} & 
		$0.9785  \pm 0.0074 $ & 
		$0.9816 \pm 0.0094$ \\
		
		$H_0 \, [\mathrm{km/s/Mpc}]$ & 
		$69.68  \pm 1.00$ & 
		$70.79\pm1.41$ \\
		
		{$\Omega_\mathrm{b} h^2$} & 
		$0.02269  \pm 0.00025 $ & 
		$0.02291  \pm 0.00036$ \\
		
		{$\Omega_\mathrm{cdm} h^2$} & 
		$0.1143\pm0.0020$ & 
		$0.1178\pm0.0039$ \\
		
		{$\tau_\mathrm{reio}$} & 
		$0.0510\pm 0.0086$ & 
		$0.0511\pm 0.0085$\\
		
		{$\mathrm{log}_{10}(z_c)$} & 
		$-$ & 
		$3.75^{+0.55}_{-0.17}$ \\
		
		{$f_\mathrm{EDE} $} & 
		$-$ & 
		$< 0.104$\\
		
		{$\theta_i$} & 
		$-$ & 
		$1.60^{+1.13}_{-0.88} $\\
		
		\hline
		
		$\Omega_\mathrm{m}$ & 
		$0.2838 \pm0.0118$ & 
		$0.2822 \pm0.0120$ \\
		
		$\sigma_8$ & 
		$0.7842 \pm 0.0087$ & 
		$0.7894\pm0.0131$ \\
		
		$S_8$ & 
		$0.763  \pm 0.022$    & 
		$0.766  \pm 0.024$ \\
		
		$r_s$ & 
		$145.76\pm 0.46$   & 
		$143.71\pm 1.84$	\\
		
		\hline
	\end{tabular} 
	\caption{Marginalized constraints on the cosmological parameters in $\Lambda$CDM and in the EDE scenario with $n=3$, as inferred from the $\rm Planck\text{-}low\ell\!+\!SPT$ data. The upper limit on $\fede$ is quoted at 95\% CL. }
	\label{table:base}
\end{table*}

We find no evidence for non-zero $\fede$ in the CMB data only analysis. We report an upper bound  $\fede<0.104\,(2\sigma)$ which is compatible with the amount of EDE required to alleviate the Hubble tension. Thus, the EDE model yields substantially higher values of the Hubble parameter, $H_0=70.79\pm1.41\kms$, being in $1.6\sigma$ agreement with the SH0ES measurements.
We emphasize that the $\rm Planck\text{-}low\ell\!+\!SPT$ data allow for somewhat larger values of $\fede$ as compared to that from the full Planck likelihood, namely $\fede<0.087\,(2\sigma)$ \cite{Hill:2020osr}.
At the same time, the EDE scenario supplies substantially low values of the late-time amplitude, $S_8=0.766\pm 0.024$, being in perfect agreement with the multiple probes of LSS. This effect is attributed to the fact that the large-angular scale Planck temperature power spectrum and SPTPol data both accommodate a low $S_8$ \cite{Chudaykin:2020acu}. On the contrary, the full Planck likelihood favours substantially higher values of $\sigma_8$ and $S_8$ which are primarily driven by overly enhanced lensing smoothing of the CMB peaks in the Planck temperature spectrum. The upward shift of these parameters makes the EDE prediction incompatible with the current LSS data as reported in Ref. \cite{Hill:2020osr}. The $\rm Planck\text{-}low\ell\!+\!SPT$ data allows one to alleviate this issue making the region of appreciably high $\fede\sim0.1$ compatible with cosmological data.

The epoch of EDE is constrained to $\logzc=3.75^{+0.55}_{-0.17}$. It reliably supports only a lower bound on $z_c$, whereas the upper tail of $\logzc$ remains largely unconstrained. The posterior distribution for $\logzc$ clearly indicates one single maximum, whereas the previous EDE studies hint at a weakly bimodal distribution for that \cite{Smith:2019ihp,Hill:2020osr,Ivanov:2020ril}. As discussed in Ref. \cite{Smith:2019ihp}, this ambiguous behaviour is driven by the Planck polarization measurements at high-$\ell$ and could simply be a noise fluctuation. We also find a much flatter distribution for the initial field displacement, namely $\theta_i=1.60^{+1.13}_{-0.88}$. The previous EDE analyses which adopted the full Planck likelihood \cite{Smith:2019ihp,Hill:2020osr,Ivanov:2020ril} reveal, on the contrary, a strong preference for a large initial field displacement. This large $\theta_i$ preference comes from a oscillatory pattern in the residuals of the TE and EE Planck spectra in the multipole range $\ell\sim30-500$ \cite{Smith:2019ihp} which is disfavored by the $\rm Planck\text{-}low\ell\!+\!SPT$ data \footnote{The ACT observations also do not detect any oscillatory feature in TE and EE measurements at intermediate scales \cite{Aiola:2020azj} thus supporting our inference. This implies that the residuals observed in Planck polarization measurements are, most likely, merely caused by systematic effects \cite{Lin:2020jcb}.}. Thus, our result validates the monomial expansion of the field potential when one employs the optimally constructed $\rm Planck\text{-}low\ell\!+\!SPT$ likelihood.



\subsubsection{$Planck\text{-}low\ell\!+\!SPT$+BOSS}
\label{subsec:boss}

\begin{figure*}[ht]
	\begin{center}
		\includegraphics[width=1\textwidth]{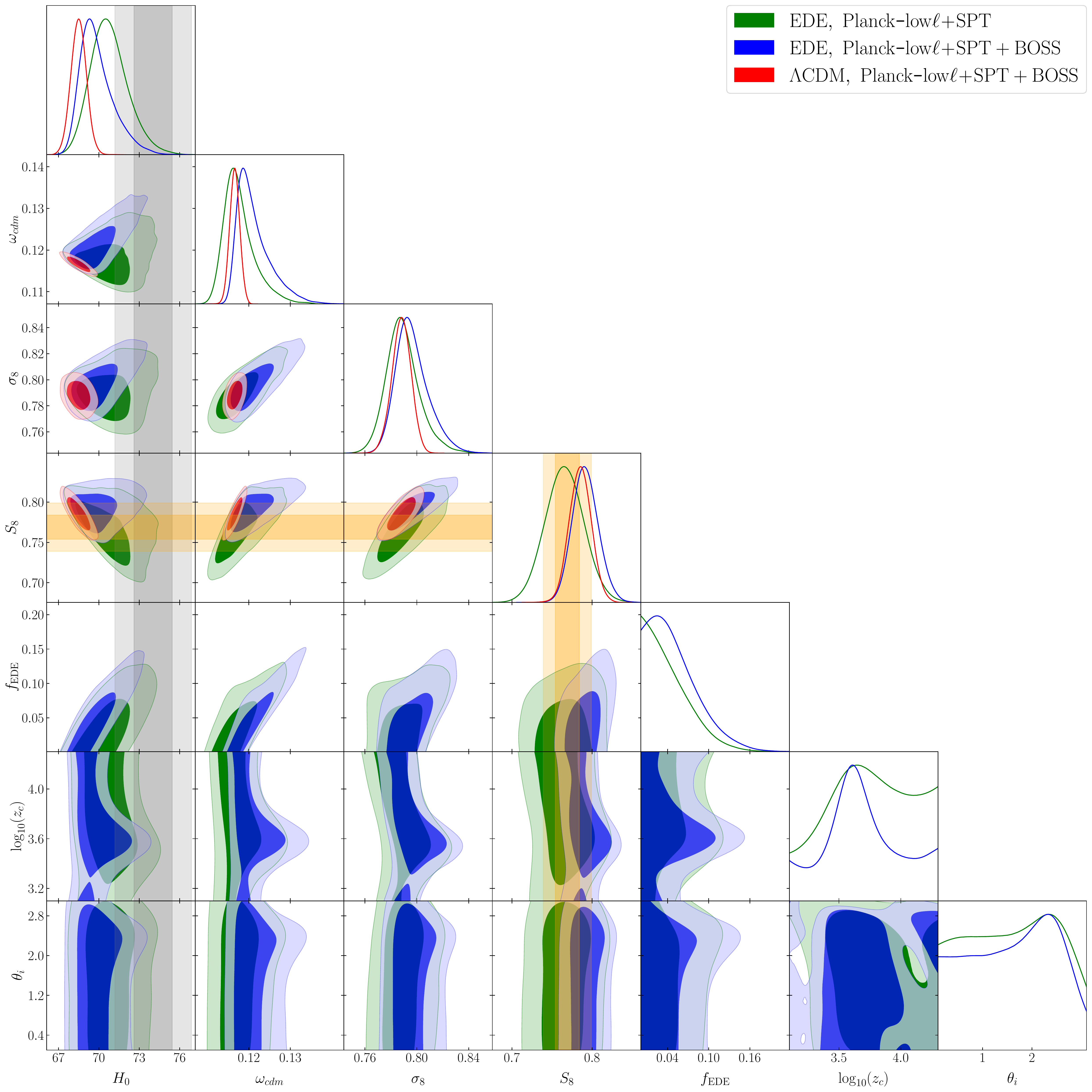}
	\end{center}
	\caption{
		Posterior distributions of the cosmological parameters of the $\Lambda$CDM and EDE model for different data sets. The gray and yellow bands represent the constraints ($1\sigma$ and $2\sigma$ confidence regions) on $H_0$ and $S_8$ coming from SH0ES and $S_8$ data, respectively.
		\label{fig:spt_1} } 
\end{figure*}

We perform a joint analysis of the $\rm Planck\text{-}low\ell\!+\!SPT$ data and the BOSS DR12 FS+BAO likelihood. The parameter constraints for the $\Lambda$CDM and EDE scenarios are presented in Tab. \ref{table:boss}. The corresponding 2d posterior distributions are shown in Fig. \ref{fig:spt_1}.

\begin{table*}[htb!]
	\renewcommand{\arraystretch}{1.1}
	Constraints from $\rm Planck\text{-}low\ell\!+\!SPT\!+\!BOSS$ \vspace{2pt} \\
	\centering
	\begin{tabular}{|l|c|c|}
		\hline Parameter &$\Lambda$CDM~~&~~~EDE \\  \hline\hline
		
    {$\ln(10^{10} A_\mathrm{s})$} & 
	$3.014 \pm 0.017$ &
	$3.019 \pm 0.018$ \\

	{$n_\mathrm{s}$} & 
	$0.9716  \pm 0.0056 $ & 
	$0.9766 \pm 0.0090$ \\

	$H_0 \, [\mathrm{km/s/Mpc}]$ & 
	$68.50  \pm 0.57$ & 
	$69.89\pm1.28$ \\

	{$\Omega_\mathrm{b} h^2$} & 
	$0.02250  \pm 0.00021 $ & 
	$0.02279  \pm 0.00039$ \\

	{$\Omega_\mathrm{cdm} h^2$} & 
	$0.1166\pm0.0012$ & 
	$0.1213\pm0.0042$ \\

	{$\tau_\mathrm{reio}$} & 
	$0.0456\pm 0.0082$ & 
	$0.0457\pm 0.0085$\\

	{$\logzc$} & 
	$-$ & 
	$3.69^{+0.61}_{-0.14}$ \\

	{$f_\mathrm{EDE} $} & 
	$-$ & 
	$< 0.118$\\

	{$\theta_i$} & 
	$-$ & 
	$1.61^{+1.13}_{-0.83} $\\

	\hline

	$\Omega_\mathrm{m}$ & 
	$0.2978\pm0.0071$ & 
	$0.2963 \pm0.0070$ \\

	$\sigma_8$ & 
	$0.7880 \pm 0.0074$ & 
	$0.7966\pm0.0129$ \\

	$S_8$ & 
	$0.785  \pm 0.014$    & 
	$0.792  \pm 0.016$ \\
	
	$r_s$ & 
	$145.31\pm 0.31$   & 
	$142.67\pm 2.14$	\\
	
		\hline
	\end{tabular} 
	\caption{Marginalized constraints (68\% CL) on the cosmological parameters in $\Lambda$CDM and in the EDE scenario with $n=3$, as inferred from the $\rm Planck\text{-}low\ell\!+\!SPT\!+\!BOSS$ data. The upper limit on $\fede$ is quoted at 95\% CL. }
	\label{table:boss}
\end{table*}

We found an appreciably weaker constraint on EDE, $\fede<0.118\,(2\sigma)$, compared to that in the analysis with CMB alone in the previous subsection. The $14\%$ alleviation of the upper bound is primarily driven by the upward shift in the mean value of $\omega_c$ needed to maintain the fit to the BOSS data. We emphasize that our analysis provides with substantially larger values of $\fede$ compared to that from the full Planck and BOSS data, $\fede<0.072\,(2\sigma)$ \cite{Ivanov:2020ril}. Despite the weaker constraint on $\fede$, the mean value of $H_0$ significantly decreases and its error bar is considerably shrunk, namely $H_0=68.52\pm 0.58\kms$ and $H_0=69.89\pm1.28\kms$ in the $\Lambda$CDM and EDE scenarios. It becomes possible due to the more precise BAO measurements which being combined with the FS information impose a tight constraint on $H_0$. Nevertheless, within the EDE scenario the Hubble tension with the SH0ES measurement is below $2.2\sigma$ statistical significance. Besides that, we found appreciably larger values of $S_8$, namely $S_8=0.785\pm 0.014$ and $S_8=0.792\pm 0.016$ in the $\Lambda$CDM and EDE models. We emphasize that these constraints are fully compatible with the galaxy clustering and weak gravitational lensing measurements that justifies further including of DES-Y1, KiDS-1000 and HSC data. Overall, the BAO+FS BOSS likelihood yields unprecedented gain of cosmological information: it provides with the two-fold improvement for the $\omega_c$ and $H_0$ measurements over those from the CMB alone.

Regarding the other EDE parameters, we found a slightly more precise measurement of the EDE epoch, $\logzc=3.69^{+0.61}_{-0.14}$. As a result, the maximum of the posterior distribution for this parameter is better visualized as can be appreciated from Fig. \ref{fig:spt_1}. We do not find any improvement for the initial field displacement, $\theta_i=1.61^{+1.13}_{-0.83}$, which still supports the substantially flat distribution. It also corroborates our CMB only analysis which does not find any evidence for subdominant peaks in the $\logzc$ distribution. This indicates that the bimodality behaviour previously claimed in the EDE analyses \cite{Smith:2019ihp,Hill:2020osr,Ivanov:2020ril} indeed comes from the Planck measurements at high-$\ell$ in full accord with the claim made in Ref. \cite{Smith:2019ihp}.

\subsubsection{$Planck\text{-}low\ell\!+\!SPT$+BOSS+$S_8$}
\label{subsec:S8}

\begin{figure*}[ht]
	\begin{center}
		\includegraphics[width=1\textwidth]{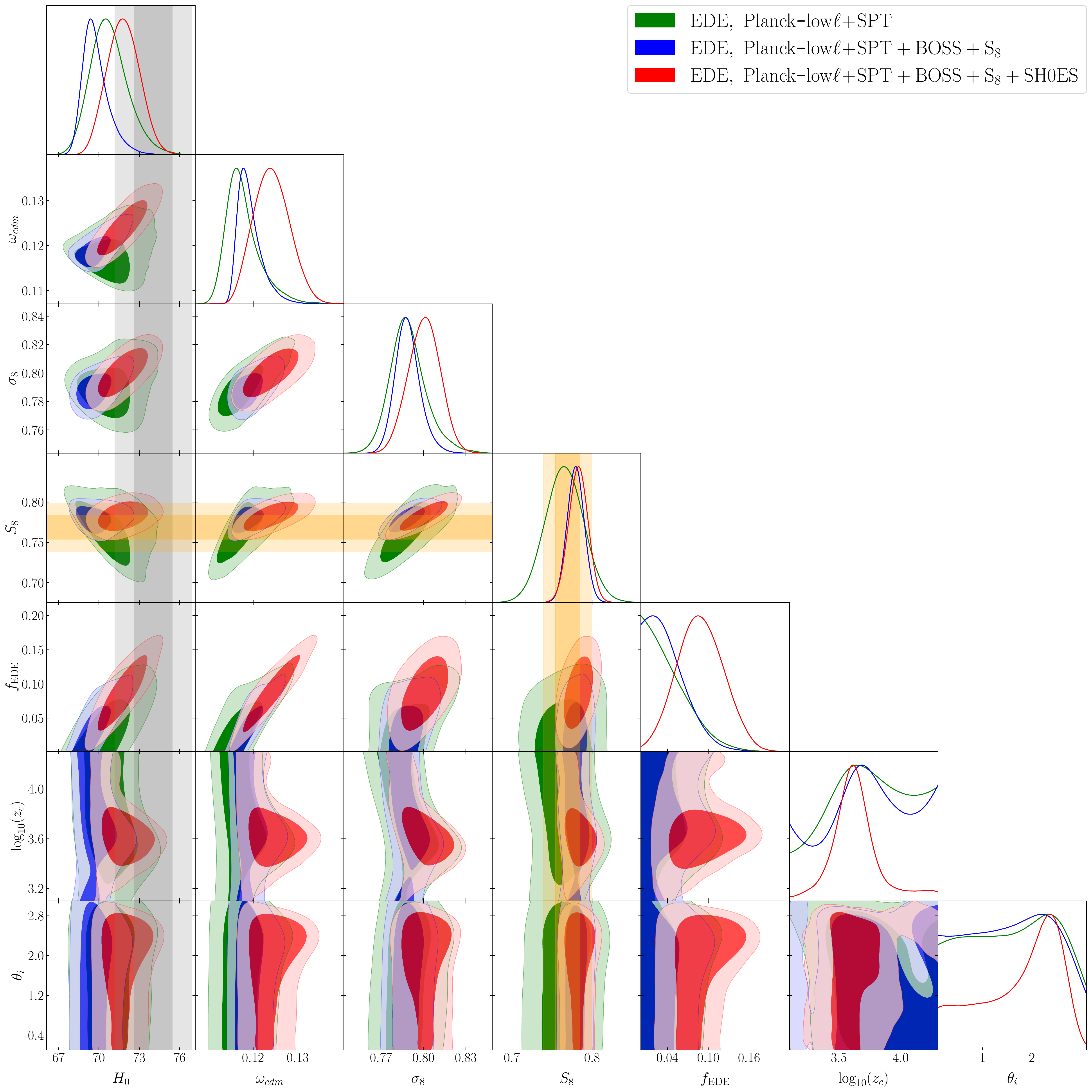}
	\end{center}
	\caption{
		Posterior distributions of the cosmological parameters of the EDE model for different data sets. The gray and yellow bands represent the constraints ($1\sigma$ and $2\sigma$ confidence regions) on $H_0$ and $S_8$ coming from SH0ES and $S_8$ data, respectively.
		\label{fig:spt_2} } 
\end{figure*}

On the next step, we include the $S_8$ information from DES-Y1, KiDS-1000 and HSC measurements by adopting the Gaussian prior $S_8=0.769\pm0.015$. The parameter constraints in the $\Lambda$CDM and EDE scenarios are reported in Tab. \ref{table:S8H0} (second and third columns). The corresponding 2d posterior distributions are shown in Fig. \ref{fig:spt_2}.

\begin{table*}[htb!]
	\renewcommand{\arraystretch}{1.1}
	Constraints from $\rm Planck\text{-}low\ell\!+\!SPT\!+\!BOSS\!+\!S_8(\!+\!SH0ES)$ \vspace{2pt} \\
	\centering
	\begin{tabular}{|l|c|c|c|}
		\hline Parameter &$\Lambda$CDM~~&~~~EDE &~~~EDE $\rm (+SH0ES)$\\  \hline\hline
		
		{$\ln(10^{10} A_\mathrm{s})$} & 
		$3.008 \pm 0.017$ &
		$3.013 \pm 0.017$ &
		$3.021 \pm 0.017$ \\
		
		{$n_\mathrm{s}$} & 
		$0.9735  \pm 0.0054 $ & 
		$0.9753 \pm 0.0076$ &
		$0.9870 \pm 0.0089$ \\
		
		$H_0 \, [\mathrm{km/s/Mpc}]$ & 
		$68.82  \pm 0.50$ & 
		$69.79\pm0.99$ &
		$71.81\pm1.19$ \\
		
		{$\Omega_\mathrm{b} h^2$} & 
		$0.02255  \pm 0.00020 $ & 
		$0.02276  \pm 0.00036$ &
		$0.02318  \pm 0.00042$ \\
		
		{$\Omega_\mathrm{cdm} h^2$} & 
		$0.1159\pm0.0010$ & 
		$0.1193\pm0.0028$ &
		$0.1241\pm0.0039$ \\
		
		{$\tau_\mathrm{reio}$} & 
		$0.0437\pm 0.0087$ & 
		$0.0446\pm 0.0086$ &
		$0.0448\pm 0.0089$ \\
		
		{$\logzc$} & 
		$-$ & 
		$3.74^{+0.56}_{-0.15}$ &
		$3.64^{+0.13}_{-0.18}$ \\
		
		{$f_\mathrm{EDE} $} & 
		$-$ & 
		$< 0.094$ &
		$0.088\pm0.034$ \\
		
		{$\theta_i$} & 
		$-$ & 
		$1.57^{+1.05}_{-0.86} $ &
		$1.79^{+1.02}_{-0.42} $ \\
		
		\hline
		
		$\Omega_\mathrm{m}$ & 
		$0.2938\pm0.0059$ & 
		$0.2930 \pm0.0059$ &
		$0.2870 \pm0.0055$ \\
		
		$\sigma_8$ & 
		$0.7839 \pm 0.0069$ & 
		$0.7889\pm0.0089$ &
		$0.8005\pm0.0111$ \\
		
		$S_8$ & 
		$0.776  \pm 0.010$    & 
		$0.780  \pm 0.011$	&
		$0.783  \pm 0.012$	\\
		
		$r_s$ & 
		$145.45\pm 0.28$   & 
		$143.51\pm 1.50$	&
		$140.61\pm 2.01$	\\
		\hline
	\end{tabular} 
	\caption{Marginalized constraints (68\% CL) on the cosmological parameters in $\Lambda$CDM and in the EDE scenario with $n=3$, as inferred from the $\rm Planck\text{-}low\ell\!+\!SPT\!+\!BOSS\!+\!S_8$ (second and third columns) and $\rm Planck\text{-}low\ell\!+\!SPT\!+\!BOSS\!+\!S_8\!+\!SH0ES$ (fourth column) data sets. The upper limit on $\fede$ is quoted at 95\% CL. }
	\label{table:S8H0}
\end{table*}

We found a more stringent limit on EDE, $\fede<0.094\,(2\sigma)$, which represents a $20\%$ improvement over that from the analysis without $S_8$ information in the previous subsection. This gain is explained by a $0.5\sigma$ downward shift in $\omega_c$ which strongly correlates with $\fede$. A lower value of $\omega_c$, in turn, reduces the growth rate of perturbations at late times that allows one to accommodate a substantially lower  value of $S_8$ in accord with the $S_8$ data. In particular, the combined $\rm Planck\text{-}low\ell\!+\!SPT\!+\!BOSS\!+\!S_8$ data bring $S_8=0.776  \pm 0.011$ and $S_8=0.780  \pm 0.011$ in the $\Lambda$CDM and EDE scenarios which represent $20\%$ and $30\%$ improvements over that in the $\rm Planck\text{-}low\ell\!+\!SPT\!+\!BOSS$ analysis (without $S_8$). More precise determination of $S_8$ improves the $H_0$ constraints to the same extent, namely $H_0=68.84  \pm 0.50\kms$ and $H_0=69.79\pm0.99\kms$ in the $\Lambda$CDM and EDE scenarios. We emphasize that the mean values of $H_0$ upon including the $S_8$ information remain essentially unchanged that demonstrates a high level of compatibility between the $\rm Planck\text{-}low\ell\!+\!SPT\!+\!BOSS$ and $S_8$ data. It is worth nothing that the $H_0$ constraint inferred in the $\Lambda$CDM scenario is in a substantial $3.5\sigma$ tension with the Cepheid calibrated local measurement of $H_0$. In the EDE model this tension is alleviated to an allowable level of $2.5\sigma$ that makes the $\rm Planck\text{-}low\ell\!+\!SPT$+BOSS+$\rm S_8$ and SH0ES data statistically compatible within the EDE framework.

Upon including the $S_8$ information we do not find a substantial improvement in the $\logzc$ and $\theta_i$ measurements. We found $\logzc=3.74^{+0.56}_{-0.15}$ and $\theta_i=1.57^{+1.05}_{-0.86}$ which are consistent with the results of the previous analysis without the $S_8$ data. The virtually intact constraints on these parameters can be readily understood. Unlike $\fede$, the $\logzc$ and $\theta_i$ very weakly correlate with the other $\Lambda$CDM parameters as demonstrated in Fig. \ref{fig:spt_1}. It implies that a more precise measurements of $S_8$ without a significant shift in its mean value can not substantially alter the posterior distributions of $\logzc$ and $\theta_i$ parameters. 

\subsubsection{$Planck\text{-}low\ell\!+\!SPT$+BOSS+$S_8$+SH0ES}
\label{subsec:H0}

We finally address the Cepheid-based local measurements of $H_0$. Since the SH0ES measurement and $\rm Planck\text{-}low\ell\!+\!SPT$+BOSS+$\rm S_8$ inference of $H_0$ are in the significant tension in the framework of $\Lambda$CDM cosmology, we do not combine them in one data set under the $\Lambda$CDM assumption. The parameter constraints for the EDE scenario are presented in Tab. \ref{table:S8H0} (fourth column). The corresponding 2d posterior distributions are shown in Fig. \ref{fig:spt_2}.

We found $\fede=0.088\pm0.034$, which indicates a $2.6\sigma$ preference for nonzero EDE. This result is driven by the SH0ES measurement which favours a substantially higher value of $H_0$. Namely, we found $H_0=71.81\pm1.19\kms$ which is now in $1.2\sigma$ agreement with the SH0ES data. We emphasize that the error bar on $H_0$ increases quite moderately, by $20\%$ over that from the analysis without SH0ES. It indicates that a better agreement with the SH0ES measurement comes from a released freedom in the EDE model rather than to be a result of the worse description of other data sets. Indeed, the constraint on $S_8$ remains virtually unchanged compared to the analysis without SH0ES, namely $S_8=0.783  \pm 0.012$. It implies that a higher $H_0$ does not significantly degrade the fit to the LSS data. Thus, our approach based on the $\rm Planck\text{-}low\ell\!+\!SPT$ CMB data alleviates the conflict between the SH0ES-tension-resolving EDE cosmologies and LSS data previously claimed in Ref. \cite{Hill:2020osr,Ivanov:2020ril}.

Addressing the local $H_0$ measurements significantly alters the posterior distribution for $\logzc$ and $\theta_i$ parameters. We found a more stringent constraint on the EDE epoch, $\logzc=3.64^{+0.13}_{-0.18}$, as opposed to the more flattened distribution observed in the previous analyses without SH0ES. This result distinctively indicates a narrow redshift interval prior to recombination at which EDE efficiently decays in full accord with the EDE proposal. We also find a strong preference for large initial field displacement, $\theta_i=1.79^{+1.02}_{-0.42}$, consistent with the findings of \cite{Smith:2019ihp,Hill:2020osr}. Our results reflect how the SH0ES measurements break parameter degeneracies in the EDE sector to restore cosmological concordance.

To scrutinize the impact of the SH0ES measurements, we examine the goodness-of-fit for each individual likelihood. For that, we provide the best-fit $\chi^2$ values for each data set optimized to the data with and without SH0ES measurements in Tab. \ref{tab:chi2}.
\begin{table}
	\renewcommand{\arraystretch}{1.0}
	\centering
	\begin{tabular} {| c | c |c | c|}
		\hline
		Data set & w/o SH0ES & w SH0ES & $N_{\rm dof}$ \\
		\hline
		\hline
		$\rm Planck\text{-}low\ell$ & $825.43$ & $825.45$ & $1005$ \\ 
		$\rm SPT$ & $148.82$ & $145.83$ & $114$ \\ 
		$\rm BOSS$ & $476.55$ & $478.80$ & $372$ \\ 
		$\rm S_8$ & $<0.01$ & $1.03$ & $1$ \\
		$\rm SH0ES$ & $10.23$ & $1.13$ & $1$ \\
		\hline
		$\sum\chi^2_{\rm -SH0ES}$ & $1450.79$ & $1451.10$ & $1492$ \\
		$\sum\chi^2$ & $1461.02$ & $1452.23$ & $1493$ \\
		\hline
	\end{tabular}
	\caption {$\chi^2$ values for the best-fit EDE model optimized to the $\rm Planck\text{-}low\ell\!+\!SPT\!+\!BOSS\!+\!S_8$ (second column) and $\rm Planck\text{-}low\ell\!+\!SPT\!+\!BOSS\!+\!S_8\!+\!SH0ES$ (third column) data.
	We also provide the number of degrees of freedom $N_{\rm dof}=N_{\rm data}-N_{\rm fit}$ where $N_{\rm data}$ is the number of data points and $N_{\rm fit}$ is the sum of fitting parameters (fourth column)}.
	\label{tab:chi2}
\end{table}
We found that the total $\chi^2$-statistics optimized to the data without SH0ES changes insignificantly, $\Delta\chi^2_{\rm -SH0ES}=0.31$. It implies that adding SH0ES does not considerably spoil the fit to the other data. 
In particular, including SH0ES does not afflict the fit to the $\rm Planck\text{-}low\ell$ likelihood. In turn, it moderately degrades the fit to the BOSS data by $\Delta\chi^2_{\rm BOSS}=2.25$ \footnote{We caution that the reduced $\chi^2$-statistics, $\chi^2/N_{\rm dof}$, is a very inaccurate metric for the goodness of fit. First, it does not accounts for the covariance between different data bins. Second, it assumes that cosmological information is uniformly distributed between different data bins which is not the case for the BOSS data. Indeed, the reduced $\chi^2$ for the BOSS likelihood reads $\chi^2/N_{\rm dof}=1.29$, see Tab. \ref{tab:chi2}, which can be naively interpreted as a bad fit. However, the reduced $\chi^2$ substantially decreases if one exploits a substantially larger momentum cutoff $k_{\min}=0.05~\hMpc$ instead of $k_{\min}=0.01~\hMpc$ used in our analysis as demonstrated in Ref. \cite{Ivanov:2019pdj}.}. This effect can be readily understood. A higher value of $H_0$ driven by SH0ES is accommodated by higher $\fede$ which, in turn, strongly correlates with $\omega_c$ and $\sigma_8$ as shown in Fig. \ref{fig:spt_2}. However, higher values of these parameters become at odds with the BOSS likelihood that favours moderately lower values of $\omega_c$ and $\sigma_8$. The worsening of the BOSS fit is entirely compensated by the improved fit to the SPT data. Namely, we found $\Delta\chi^2_{\rm Pol}=-2.24$ and $\Delta\chi^2_{\rm Lens}=-0.75$ for polarization and gravitational lensing measurements, respectively. These improvements can be attributed to considerably higher values of $H_0$ inferred from the SPTPol survey \cite{Henning:2017nuy,Bianchini:2019vxp}. Finally, the fit to the $S_8$ data degrades quite marginally, $\Delta\chi^2_{\rm S_8}=1.03$.  It demonstrates that the EDE scenario can accommodate a higher value of $H_0$ while not significantly deteriorating the fit to the galaxy clustering and weak lensing measurements. 

To reliably predict the preference of the EDE scenario over $\Lambda$CDM we resort to several statistical tools. First, we employ an essentially frequentist Akaike Information Criteria (AIC) \cite{1100705} that sets the penalty for extra free parameters in more complex models. For that, we build up the absolute difference in logarithmic likelihoods $\log L$ calculated for EDE and $\Lambda$CDM models at their respective best-fit points optimized to the $\rm Planck\text{-}low\ell\!+\!SPT\!+\!BOSS+\!S_8\!+\!SH0ES$ likelihood. The AIC states that the quantity $2\Delta\log L$ defined in this way is distributed as $\chi^2_n$ with $n$ degrees of freedom equal to the difference of fitting parameters in $\Lambda$CDM and EDE models. As the EDE model has three extra parameters, we put $n=3$. We found $2\Delta\log L=9.3$ that indicates a quite moderate $2.2\sigma$ preference for the EDE scenario over $\Lambda$CDM. Second, we apply a more sophisticated Bayesian evidence analysis which is ought to be preferred in model comparison since it addresses the prior volume effects that allows one to directly control the lack of predictivity of more complicated models \cite{zbMATH03189754}. The AIC does not account for the prior information which is highly relevant for model comparison questions and omitting it would result in seriously wrong inferences \cite{Trotta:2008qt}. To avoid this, we employ the \texttt{MCEvidence} code \cite{Heavens:2017afc} to estimate the Bayesian factor, the ratio between EDE and $\Lambda$CDM evidences, 
\begin{equation}
B=\frac{p({\rm EDE}|d)}{p({\rm \Lambda CDM}|d)}.
\end{equation}
Using the $\rm Planck\text{-}low\ell\!+\!SPT\!+\!BOSS+\!S_8\!+\!SH0ES$ data set we found $B=1.8$ that supports the EDE preference over $\Lambda$CDM. This preference is rather weak according to Jeffrey’s scale \cite{zbMATH03189754} due to a significantly larger parameter space volume in the EDE model compared to that in $\Lambda$CDM. Our result reveals an importance of the Bayesian evidence that penalizes models with a large volume of unconstrained parameter space more harshly than the AIC \cite{Trotta:2008qt}.



\section{Conclusions}
\label{sec:conc}

The EDE scenario is a compelling early-time resolution of the persistent and increasingly significant tension between local and global measurements of the Hubble constant. The EDE model successfully decreases the sound horizon enabling a higher value of $H_0$ in concordance with the SH0ES measurement. Accompanying shifts in $\omega_c$ and $n_s$ parameters produce the CMB power spectra nearly indistinguishable from that in the $\Lambda$CDM scenario, hence providing a good fit to both the primary CMB and distance-ladder $H_0$ data. However, as demonstrated in Ref. \cite{Hill:2020osr}, the shifts in the standard $\Lambda$CDM parameters are in tension with the various LSS data, in particular measurements of galaxy clustering and weak gravitational lensing. The region of parameter space capable of addressing the Hubble tension is further constrained when the full BOSS galaxy power spectrum likelihood is included \cite{Ivanov:2020ril}. In this paper, we revisit these stringent limits on EDE using a different CMB setup.

In fact, past claims of tight constraints on the EDE scenario \cite{Hill:2020osr,Ivanov:2020ril} essentially rely on the full Planck data. However, the Planck's residuals of the CMB temperature power spectrum present a curious oscillatory feature conventionally attributed to the extra smoothing effect of gravitational lensing that pulls the late-time amplitude to a higher value \cite{Addison:2015wyg,Aghanim:2018eyx}. Although the lensing-like anomaly does not significantly alter the $\Lambda$CDM predictions \cite{Aghanim:2016sns,Aghanim:2018eyx}, its effect may be crucial for the various extensions of the $\Lambda$CDM model which open up a new degeneracy direction with $\sigma_8$. This is indeed the case for the EDE scenario which dictates a higher value of $\sigma_8$ for the SH0ES-tension-resolving cosmology due to the tight correlation between $\sigma_8$, $\fede$ and $H_0$ parameters.
The overly enhanced smoothing of CMB peaks in the Planck temperature power spectrum increases $\sigma_8$ even further that exacerbates the discrepancy between the Planck and LSS data in the EDE framework.
As demonstrated in Ref. \cite{Murgia:2020ryi}, the full Planck likelihood indeed provides the more stringent constraints on EDE compared to the 'unlensed' CMB power spectra. Since we do not know what is behind the Planck lensing anomaly, the more conservative approach would be to analyse the CMB data without this feature. The one way is to marginalize over the lensing information in the Planck CMB power spectra as done in Ref. \cite{Murgia:2020ryi}. The second approach refers to the usage of alternative CMB measurements. In our analysis we employ the second treatment and combine the Planck and SPTPol measurements following the strategy of Ref. \cite{Chudaykin:2020acu}.

In this work, we reanalyse the EDE scenario using the Planck and SPTPol measurements of the CMB anisotropy, the full BOSS likelihood and photometric galaxy clustering and weak lensing data. As the primary CMB data we consider the Planck TT power spectrum at $\ell<1000$ and the SPTPol measurements of TE, EE and lensing potential power spectra. It has been shown in Ref. \cite{Chudaykin:2020acu} that this CMB setup predicts the consistent CMB lensing effect in the $\Lambda$CDM cosmology which modelling is important for resulting EDE constraints \cite{Murgia:2020ryi}. In this paper, we extend the previous EDE analysis \cite{Chudaykin:2020acu} assuming a more motivated power-law cosine potential \eqref{PoulinEDE} and implementing a full perturbative dynamics of the EDE field. 

We find no evidence for EDE in the primary CMB data alone: the fit to $\rm Planck\text{-}low\ell\!+\!SPT$ brings $\fede<0.104\,(2\sigma)$. Our CMB analysis yields considerably higher values of the Hubble parameter, $H_0=70.79\pm1.41$, albeit with  a $40\%$ larger error bar compared to that from the full Planck data \cite{Hill:2020osr}. Upon including the full BOSS galaxy power spectrum likelihood we found a somewhat looser constraint on EDE, $\fede<0.118\,(2\sigma)$. The mean value of the Hubble constant is shifted considerably downwards, $H_0=69.89\pm1.28\kms$. Supplemented with additional LSS data in the form of a Gaussian prior on $S_8$ from DES-Y1 \cite{Abbott:2017wau}, KiDS \cite{Asgari:2020wuj} and HSC \cite{Hikage:2018qbn} photometric measurements (a procedure was validated for the EDE model in \cite{Hill:2020osr}), we obtain a considerably tighter constraint on EDE, $\fede<0.094\,(2\sigma)$ and $H_0=69.79\pm0.99\kms$. We emphasize that even after taking into account the data from photometric surveys the available $\fede$ values are still capable of addressing the Hubble tension in contrast to the past EDE analyses which fail to simultaneously resolve the Hubble tension and maintain a good fit to both CMB and LSS data \cite{Hill:2020osr,Ivanov:2020ril}. The main culprit of the past strong constraints on EDE is the overly enhanced smoothing effect of acoustic peaks that mainly affects the Planck temperature spectrum at high-$\ell$ and pulls $\sigma_8$ to a higher value thereby conflicting with the LSS constraints. We also found that the $H_0$-tension with the SH0ES measurements is alleviated to an acceptable $2.5\sigma$ level that enables one to include the SH0ES data in the fit.

Finally, we fit the EDE model to the combined data set with SH0ES. We found a $2.6\sigma$ evidence for non-zero EDE, $\fede=0.088\pm0.034$. Our measurements reconcile the tension with the SH0ES-only constraint leading to $H_0=71.81\pm1.19\kms$. We emphasize that our inference of the Hubble constant yields a considerably higher mean value with only modestly larger error bar compared to the results of past EDE studies that utilize a similar combination of data sets (with the high-$\ell$ Planck data) \cite{Poulin:2018cxd,Smith:2019ihp,Hill:2020osr}. We scrutinize the impact of the SH0ES data on the goodness-of-fit of each individual measurement. We found that the inclusion of SH0ES moderately degrades the fit to the BOSS likelihood but this effect is entirely compensated by the improved fit to the SPTPol data. At the same time, the goodness-of-fit of the photometric LSS data is only marginally deteriorated upon including SH0ES in the analysis. It reconciles the conflict between the SH0ES-tension-resolving EDE cosmologies and LSS data previously claimed in Ref. \cite{Hill:2020osr,Ivanov:2020ril}. The AIC criteria indicates a mild preference for the EDE scenario over $\Lambda$CDM at a $2.2\sigma$ level. The statistical analysis based on the Bayesian evidence reveals an even weaker evidence for the EDE scenario. It is caused by a significantly larger volume of parameter space available in the EDE model.

Overall, our results indicate that the combined analysis of Planck and SPTPol data can address a higher $H_0$ value in full concordance with the SH0ES measurement whilst not substantially worsening the fit to the galaxy clustering and weak lensing measurements. This inference is mainly driven by two causes. First, our primary CMB data perfectly agree with the various LSS data within the $\Lambda$CDM cosmology \cite{Chudaykin:2020acu} that opens up a new region of higher $\sigma_8$ values which can accommodate a higher $H_0$ in the EDE scenario. Second, the combined data approach provides considerably larger error bars as compared to those in the baseline Planck analysis \cite{Aghanim:2018eyx} that facilitates a resolution of the Hubble tension. In particular, the fit to $\rm Planck\text{-}low\ell\!+\!SPT$ yields the $H_0$ measurement with twice the error bar and a $40\%$ looser constraint on $S_8$ \cite{Chudaykin:2020acu} as compared to that in the full Planck data analysis assuming the $\Lambda$CDM cosmology. All this makes the LSS constraints on EDE rather harmless. The constraints obtained in this work are similar to those from Ref. \cite{Murgia:2020ryi} that examines the lensing-marginalized Planck power spectra.

Likewise, this paper underlines the extreme importance of CMB lensing effect for obtaining solid constraints on EDE. Future high-resolution Simons Observatory’s Large Aperture Telescope (CMB-S4) \cite{Ade:2018sbj} will probe the CMB anisotropy with pinpoint accuracy thus providing the robust measurement of the CMB lensing effect down to very small scales. Upgrading the ongoing ground-based experiments such as SPT-3G \cite{Benson:2014qhw,Anderson:2018mry} and AdvACTPol \cite{Calabrese:2014gwa,Li:2018uwb} will continue to make progress towards more precise measurements of the CMB anisotropies on very small scales. The complementary information delivered by Cosmology Large Angular Scale Survey (CLASS) \cite{Essinger-Hileman:2014pja,Xu:2019rne} and Simons Array \cite{POLAR,Faundez:2019lmz} can also shed light on the gravitational lensing effect. Future progress provided by upcoming and ongoing CMB measurements will most probably make clear whether the lensing-like anomaly in the Planck data is real or it is merely driven by systematic effects.

\vspace{1cm}
\section*{Acknowledgments}

We  are  indebted  to  Mikhail  M.  Ivanov for his collaboration on the initial stages of this project. The work is supported by the RSF grant 17-12-01547. All numerical calculations have been performed with the HybriLIT heterogeneous computing platform (LIT, JINR) (\href{http://hlit.jinr.ru}{http://hlit.jinr.ru}).

\appendix 

\section{Nuisance parameters and priors}
\label{nuisance}

In this Appendix we describe the nuisance parameters and their priors adopted in the full-shape power spectrum analysis. For each data chunk (high-z NGC, high-z SGC, low-z NGC and low-z SGC) of the BOSS galaxy sample we independently vary the following nuisance parameters 
\be 
(b_1,b_2,b_{\mathcal{G}_2},b_{\Gamma_3},c_0,c_2,\tilde{c},P_{\rm shot}),
\ee 
where $b_1$ is the linear galaxy bias, $b_2$ is the density quadratic bias, $b_{\mathcal{G}_2}$ and $b_{\Gamma_3}$  are the quadratic and cubic
tidal biases, $c_0$ and $c_2$ are the higher-derivative $k^2$ counterterms for the monopole and quadrupole, $\tilde{c}$ is the next-to-leading order $k^4$ redshift-space counterterm, $P_{\rm shot}$ is the constant shot noise contribution. We adopt the following priors on these nuisance parameters
\be 
\label{pr}
\begin{split}
& b_1\in \text{flat}[1,4],\quad
b_2\sim \mathcal{N}(0,1^2),\\
& 
b_{\mathcal{G}_2}\sim \mathcal{N}(0,1^2),
\quad b_{\Gamma_3}\sim \mathcal{N}(0.65,1^2), \\
& c_0\sim \mathcal{N}(0,30^2),\quad 
c_2\sim \mathcal{N}(30,30^2)\\
& \tilde{c}
\sim \mathcal{N}(500,500^2),\quad 
P_{\rm shot}
\sim 
\mathcal{N}(0,5\cdot 10^3)\,,
\end{split}
\ee 
where $\mathcal{N}(a,\sigma^2)$ denotes the Gaussian distribution with the mean $a$ and standard deviation $\sigma$. 

In the following we outline the physical motivation behind the priors \eqref{pr}. The detailed information about the nuisance parameters and complete theoretical model can be found in Ref. \cite{Chudaykin:2020aoj}. The mean values of $b_2$ and $b_{\mathcal{G}_2}$ come from N-body simulations \cite{Lazeyras:2017hxw}. The mean value of $b_{\Gamma_3}$ is motivated by the coevolution model \cite{Desjacques:2016bnm}. On the EFT grounds all bias parameters are expected to be $\mathcal{O}(1)$ that motivates our priors \eqref{pr}. On general grounds, the higher-derivative counterterms, $c_0$ and $c_2$, are expected to be $\mathcal{O}(1)\times k_{\rm NL}^{-2}$ where $k_{\rm NL}$ denotes the non-linear scale. However, the non-linear scale in redshift space depends on the velocity dispersion of the BOSS galaxies, which is rather high, $\sigma_v\sim 5\,{\rm Mpc}/h$ \cite{Beutler:2016arn}. Since it is several times larger than the real-space estimate $k_{\rm NL}^{-1}\sim2\,{\rm Mpc}/h$, we approximate $c_2\sim\sigma_v^2\sim 30\,[{\rm Mpc}/h]^2$. The next-to-leading order redshift-space conterterm, $\tilde{c}$, is also dominated by strong fingers-of-God effects found in the BOSS galaxy sample that imply $\tilde{c}\sim\sigma_v^4\sim500\,[{\rm Mpc}/h]^4$. We assume the wide enough priors on $c_0$, $c_2$ and $\tilde{c}$ to accommodate a large velocity dispersion that explains our choice of \eqref{pr}. The shot noise contribution $P_{\rm shot}$ is expected to deviate from the Poissonian prediction due to exclusion effects and fiber collision \cite{Chudaykin:2020aoj}. To accommodate these effects we impose a conservative large prior centered at this bestfit value \eqref{pr}.

In general, our treatment of the short-scale physics and galaxy bias is very agnostic because we marginalize over unknown values and time-dependence of all nuisance parameters. The physical priors \eqref{pr} leads only to a better convergence of the MCMC chains and do not substantially affect the resulting posterior distribution \cite{Chudaykin:2020aoj}. The one-loop modelling of the galaxy power spectrum has been verified against multiple high-resolution mock data with a large simulation volume which demonstrates the high performance and robustness of the EFT of LSS approach \cite{Nishimichi:2020tvu,Chudaykin:2020hbf}.

\bibliographystyle{JHEP}
\bibliography{short.bib}

\end{document}